# Liquid Xenon for a Very Sensitive WIMP Search[*]

David B. Cline

*Department of Physics and Astronomy, Box 951547*
*University of California, Los Angeles, CA 90095-1547 USA*

**Abstract.** We describe the research and development efforts of the UCLA‑Torino group to develop a large, powerful, discriminating WIMP detector. We also discuss the possible amplifications of the WIMP signal. The first real detector, ZEPLIN II, is being prepared for the UKDMC Boulby Mine Laboratory.[1-3]

## STATUS OF THE SEARCH FOR WIMPs

The current status of the search for WIMPs is confused:
1. The level is at about 1/2 event/kg/d;
2. The DAMA group is making strong claims for a discovery;
3. The CDMS group has shown that their data and those of DAMA are incompatible to 99.5% confidence level against the observation of WIMPs.

Theory, which allows WIMP rates from about $10^{-1}$ to $10^{-4}$ events/kg/d, gives poor guidance.[4,5] We believe that it is essential that one method (A) resolve the CDMS/DAMA claims,[6] and (B) cover the entire region of $10^{-1}$ to $10^{-4}$ events/kg/d. The detectors described here (ZEPLIN II and IV)[1-3] can carry out this search. In Fig. 1, we show the different methods of discrimination against background.

## Properties of Liquid Xenon and Development of a Xenon WIMP Detector

The key properties of liquid Xe are given in Table 1, and Fig. 2 shows the key method of discrimination.[1,2]

Starting in the early 1990s, the UCLA‑Torino ICARUS group initiated the study of liquid Xe as a WIMP detector with powerful discrimination. Our most recent effort is the development of the two-phase detector. Figure 3 shows our 1-kg, two-phase detector and the principle of its operation. WIMP interactions are clearly discriminated from all important background by the amount of free electrons that are drifted out of the detector into the gas phase where amplification occurs. In Fig. 4, we show the resulting separation between backgrounds and simulated WIMP interactions (by neutron interaction). It is obvious from this plot that the discrimination is very powerful.

---

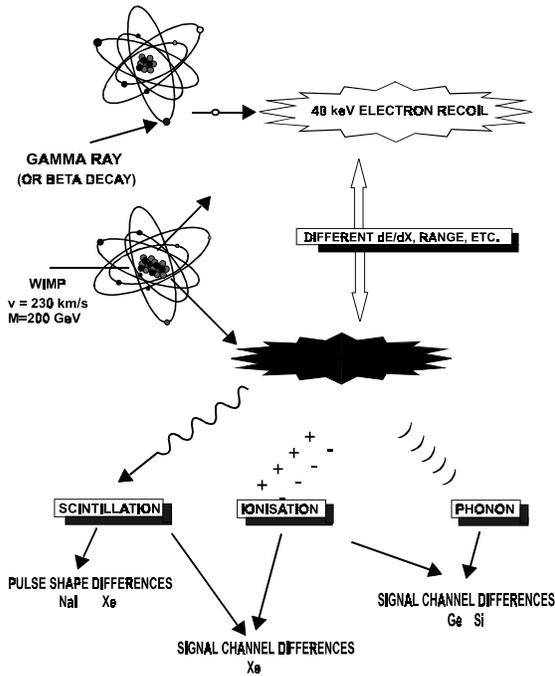 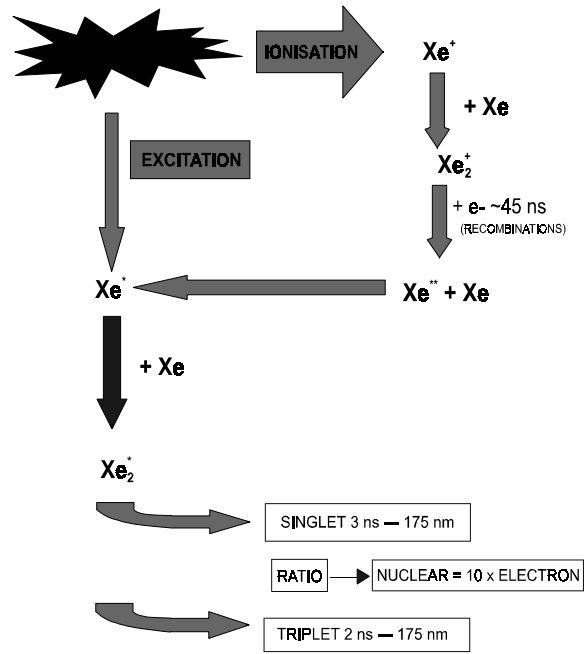

**FIGURE 1.** Methods of background discrimination for various WIMP detectors.

**FIGURE 2.** Basic mechanism for the signal and background detection in liquid Xe.

**TABLE I. Liquid Xenon as a WIMP Detector.**
_________________________________________________
1. Large mass available - up to tons.
   - Atomic mass: 131.29
   - Density: 3.057 gm/cm$^3$
   - $W_i$ value (eV/pair) 15.6 eV
   - No long-lived isotopes of xenon

2. Drift velocity: 1.7 mm/μs
   - 250 V/cm
   - Scintillation wave length: 175 nm
   - Decay time: 2 ns → 27 ns

3. Light yield > NaI, but intrinsic scintillator (no doping)

⇒ Excimer process very sell understood
⇒ First excimer laser was liquid xenon in 1970!
_________________________________________________

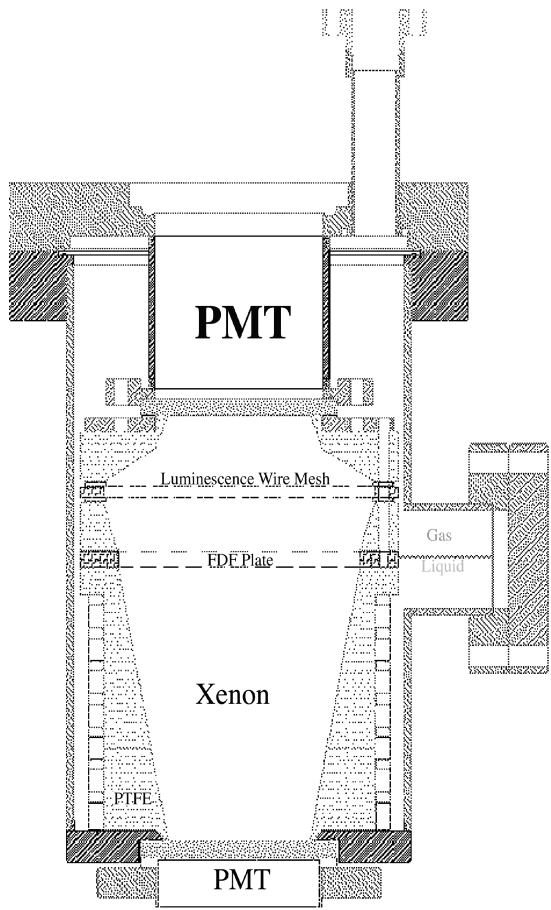
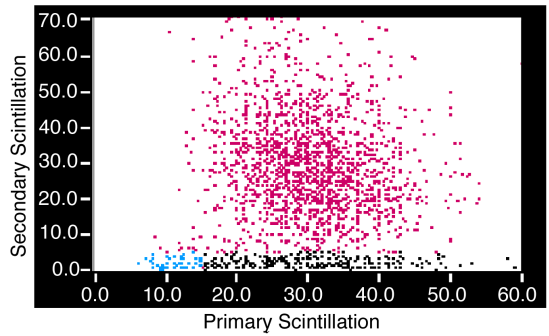

**(A)**

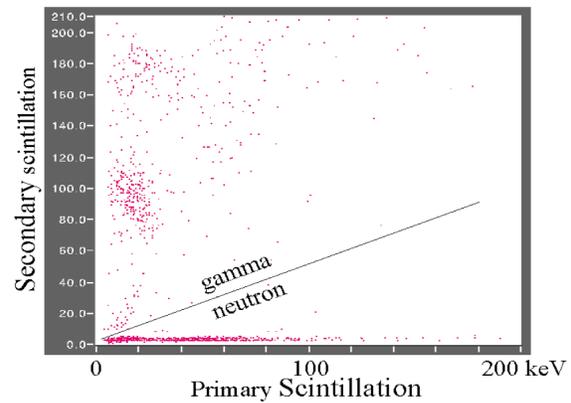

**(B)**

**FIGURE 3.** XEPLIN II: Electroluminescence in gas (principle of a two-phase, 1-kg detector, developed by UCLA‑CERN‑Torino). photons

**FIGURE 4.** Secondary vs primary scintillation plot in pure liquid Xe with mixed gamma-ray and neutron sources. The secondary scintillation are produced by (A) proportional scintillation process in liquid Xe and (B) electroluminescent process in gaseous Xe.

## ZEPLIN II DESIGN AND CONSTRUCTION

After the success of the 1-kg, two-phase detector, two directions are being taken:
(1) Construction of a large two-phase detector to search for WIMPs. The UCLA‑Torino group has formed a collaboration with the UK Dark Matter team to construct a 40-kg detector (ZEPLIN II) for the Boulby Mine underground laboratory (Fig. 5).
(2) Continuation of the R&D effort with liquid Xe to attempt to amplify the very weak WIMP signal. This work will constitute the PhD thesis of J. Woo (UCLA). The first idea to test is inserting a CsI internal photo cathode to convert UV photons to electrons that are subsequently amplified by the gas phase of the detector. In Fig. 5 we show the latest design of the ZEPLIN II detector.

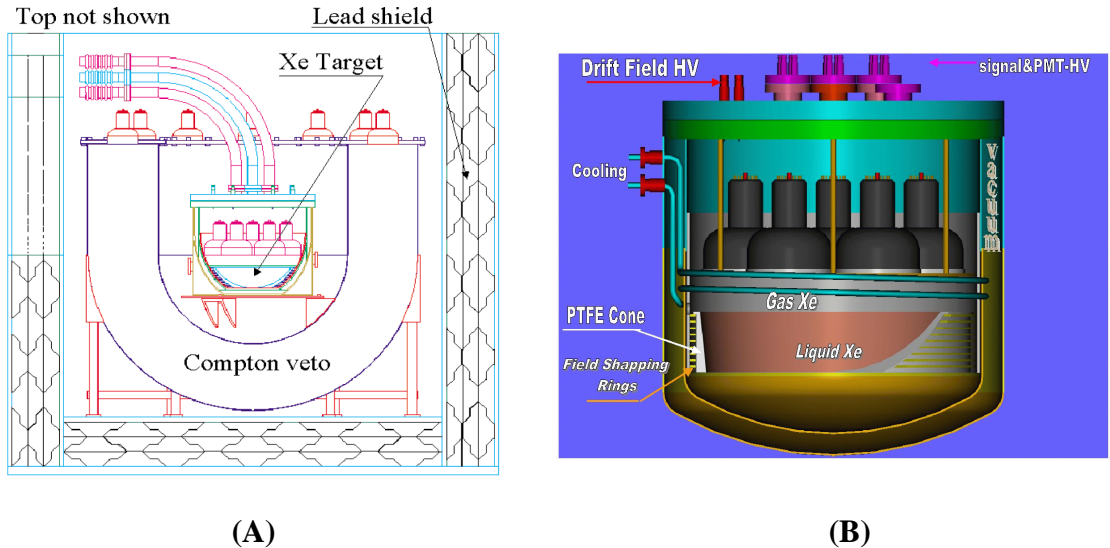

**FIGURE 5.** System setup for Xe target (40 kg total): (A) Overall set up and (B) ZEPLIN II central detector.

## POSSIBLE FUTURE LARGER XENON DETECTOR (ZEPLIN IV)

In the future we plan to design a 500-kg detector, ZEPLIN IV. We believe this detector can be the most sensitive in the world. In Fig. 6 we show the limits that ZEPLIN II and ZEPLIN IV can reach in the WIMP search.

## ACKNOWLEDGMENTS

I wish to acknowledge the excellent work of H. Wang and P. Picchi, and of the whole ICARUS team as well. In addition, I have benefitted from interactions with P. F. Smith, N. Smith, and N. Spooner and the rest of the UK Dark Matter team.

## REFERENCES

1. Cline, D., Curioni, A., Lamarina, A., et al., Astropart. Phys. **12**, 373-377 (1999).
2. Wang, H., PhD thesis, Dept. of Phys. and Astron., UCLA (1998).
3. Benetti, P. et al., Nucl. Instrum. Methods A **329**, 361 (1993).
4. Bottino, A. et al., Astropart. Phys. **2**, 77-90 (1994).
5. Nath, P. and Arnowitt, R., Phys. Rev. Lett. **74**, 4592-4595 (1995).
6. See *Sources of Dark Matter and Dark Energy in the Universe* (Proc., 4[th] Int. Symp., Marina del Rey, CA, Feb. 2000) edited by David B. Cline (Springer Verlag, Heidelberg, in press).

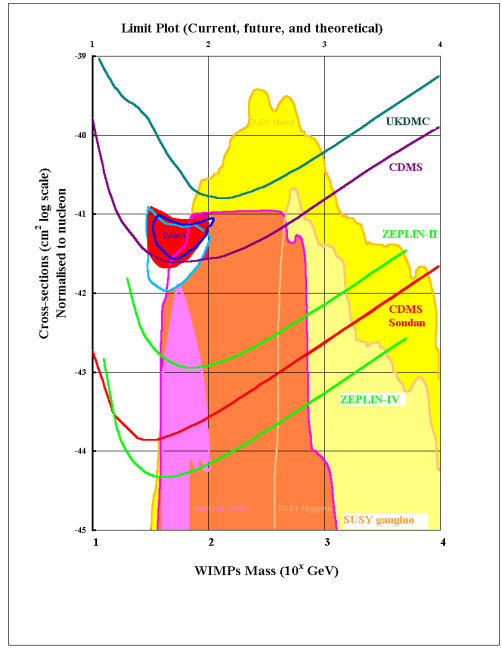

**FIGURE 6.** Limit plot (current, future, and theoretical).